# Darwin – A Mission to Detect, and Search for Life on, Extrasolar Planets

Charles S Cockell[1]
CEPSAR, The Open University, Milton Keynes, MK7 6AA, UK.
Tel : 01908 652588. Email: c.s.cockell@open.ac.uk

Alain Léger
IAS, bat 121, Universite Paris-Sud, F-91405, Paris, France

Malcolm Fridlund
Astrophysics Mission Division, European Space Agency, ESTEC, SCI-SA PO Box 299, Keplerlaan 1 NL 2200AG, Noordwijk, Netherlands.

Tom Herbst
Max-Planck-Institut fuer Astronomie, Koenigstuhl 17, 69117 Heidelberg, Germany

Lisa Kaltenegger
Harvard-Smithsonian Center for Astrophysics. 60 Garden St. MS20, Cambridge , MA 02138, USA

Olivier Absil
Laboratoire d'Astrophysique de Grenoble, CNRS, Université Joseph Fourier, UMR 5571, BP53, F-38041 Grenoble, France

Charles Beichman
Michelson Science Center, California Inst. Of Technology, Pasedena, CA 91125, USA

Willy Benz
Physikalisches Institut, University of Berne, Switzerland

Michel Blanc
Observatoire Midi-Pyrénées, 14, Av. E. Belin, Toulouse, France

Andre Brack
Centre de Biophysique Moleculaire, CNPS, Rue Charles Sadron, 45071 Orleans cedex 2, France

Alain Chelli
Laboratoire d'Astrophysique de Grenoble, CNRS, Université Joseph Fourier, UMR 5571, BP53, F-38041 Grenoble, France

Luigi Colangeli
INAF - Osservatorio Astronoomico di Capodimonte, Via Moiariello 16, 80131 Napoli, Italy

Hervé Cottin
Laboratoire Interuniversitaire des Systèmes Atmosphériques Universités Paris 12, Paris 7, CNRS UMR 7583 91, av. Di Général de Gaulle, 94010 Créteil Cedex, France

Vincent Coudé du Foresto
LESIA - Observatoire de Paris, 5 place Jules Janssen, F-92190 Meudon, France

William Danchi
Astrophysics Science Division, NASA Goddard Space Flight Center, Greenbelt, MD 20771, USA

Denis Defrère
Institut d'Astrophysique et de Géophysique de Liège, 17 Allée du 6 Août, 4000 Liège, Belgium

Jan-Willem den Herder
SRON Netherlands Institute for Space Research, Sorbonnelaan 2, 3584 CA Utrecht, Netherlands

Carlos Eiroa
Dpto Fisica Toerica C-XI, Facultad de Ciencas, Universidad Autonoma de Madrid, Cantoblanco, 28049 Madrid, Spain

Jane Greaves
University of St Andrews - Physics & Astronomy, North Haugh, St Andrews, Fife KY16 9SS, UK

Thomas Henning
Max-Planck-Institut fuer Astronomie, Koenigstuhl 17, 69117 Heidelberg, Germany

Kenneth Johnston
United States Naval Observatory, 3450 Massachusetts Avenue NW, Washington. D.C. 20392, USA

Hugh Jones
Centre for Astrophysics Research, University of Hertfordshire, College Lane, Hatfield AL10 9AB, UK

Lucas Labadie
Max Planck Institute fur Astronomie, Konigstuhl,17, 69117 Heidelberg, Germany

Helmut Lammer
Space Research Institute, Austrian Academy of Sciences, Schmiedlstr. 6, A-8042, Graz, Austria

Ralf Launhardt
Max-Planck-Institut fuer Astronomie, Koenigstuhl 17, 69117 Heidelberg, Germany

Peter Lawson
Jet Propulsion Laboratory, California Institute of Technology, 4800 Oak Grove Drive, Pasadena 91109, USA

Oliver P. Lay

Jet Propulsion Laboratory, California Institute of Technology, 4800 Oak Grove Drive, Pasedena 91109, USA

Jean-Michel LeDuigou
Centre National d'Etudes Spatiales, Optical Department,18 av. E. Belin, 31401 Toulouse cedex 9, France

René Liseau
Onsala Space Observatory, Chalmers University of Technology, SE-439 92 Onsala, Sweden

Fabien Malbet
Laboratoire d'Astrophysique de Grenoble, CNRS, Université Joseph Fourier, UMR 5571, BP53, F-38041 Grenoble, France

Stefan R. Martin
Jet Propulsion Laboratory, California Institute of Technology, 4800 Oak Grove Drive, Pasadena CA 91109, USA

Dimitri Mawet
Jet Propulsion Laboratory, California Institute of Technology, 4800 Oak Grove Drive, Pasadena CA91109, USA

Denis Mourard
Observatoire de la Côte d'Azur, Avenue Copernic, 06130 Grasse, France

Claire Moutou
Laboratoire d'Astrophysique de Marseille (LAM), CNRS, Traverse du Siphon, BP 8, Les Trois Lucs, 13376 Marseille cedex 12, France

Laurent Mugnier
ONERA/DOTA, B.P. 72, F-92322 Châtillon cedex, France

Francesco Paresce
IASF-Bologna, INAF, Italy

Andreas Quirrenbach
ZAH, Landerssternwarte, Koenigstuhl, D-69117 Heidelberg, Germany

Yves Rabbia
Observatoire de la Cote d'Azur, Dpt GEMINI   UMR CNRS 6203, Av Copernic, 06130 Grasse, France

John A. Raven
Division of Plant Sciences, University of Dundee at SCRI, Scottish Crop Research Institute, Invergowrie, Dundee DD2 5DA, UK

Huub J.A. Rottgering
Leiden Observatory, Leiden University, PO Box 9513, 2300 RA Leiden, The Netherlands

Daniel Rouan
LESIA - PHASE - Observatoire de Paris, 5 place Jules Janssen, F-92190 Meudon, France

Nuno Santos
Centro de Astrofisica, Universidade do Porto, Rua das Estrelas, 4150-762 Porto, Portugal

Franck Selsis
CRAL (CNRS UMR 5574), Université de Lyon, Ecole Supérieure de Lyon, 46 Allée d'Italie
F-69007 Lyon, France

Eugene Serabyn
Jet Propulsion Laboratory, California Institute of Technology, 4800 Oak Grove Drive,
Pasadena CA 91109, USA

Hiroshi Shibai
Graduate School of Science, Nagoya University, Furo-cho, Chikusa-ku, Nagoya
464-8602, Japan

Motohide Tamura
National Astronomical Observatory, Osawa 2-43-5, Mitaka, Tokyo 181-8588, Japan

Eric Thiébaut
Université Lyon 1, Villeurbanne, Centre de Recherche Astronomique de Lyon, Observatoire
de Lyon, CNRS/UMR-5574, Ecole Normale Supérieure de Lyon, France

Frances Westall
Centre de Biophysique Moléculaire, CNRS, Rue Charles Sadron, 45071 Orléans cedex 2,
France

Glenn J. White
Dept of Physics and Astronomy, The Open University, Walton Hall, Milton Keynes, MK7
6AA, UK
and Space Science and Technology Department, CCLRC Rutherford Appleton Laboratory,
Chilton, Didcot, Oxfordshire, OX11 0QX, UK

**ABSTRACT** - The discovery of extra-solar planets is one of the greatest achievements of modern astronomy. The detection of planets with a wide range of masses demonstrates that extra-solar planets of low mass exist. In this paper we describe a mission, called *Darwin*, whose primary goal is the search for, and characterization of, terrestrial extrasolar planets and the search for life. Accomplishing the mission objectives will require collaborative science across disciplines including astrophysics, planetary sciences, chemistry and microbiology. *Darwin* is designed to detect and perform spectroscopic analysis of rocky planets similar to the Earth at mid-infrared wavelengths (6 to 20 µm), where an advantageous contrast ratio between star and planet occurs. The baseline mission lasts 5 years and consists of approximately 200 individual target stars. Among these, 25 to 50 planetary systems can be studied spectroscopically, searching for gases such as $CO_2$, $H_2O$, $CH_4$ and $O_3$. Many of the key technologies required for the construction of *Darwin* have already been demonstrated and the remainder are estimated to be mature in the near future. *Darwin* is a mission that will ignite intense interest in both the research community and the wider public.

**INTRODUCTION**

Imaginative thoughts of worlds other than our own, perhaps inhabited by exotic creatures, have been an integral part of our history and culture. Some of the great intellects of classic civilization, such as Democritos of Abdera (460-371 BC), Epicurus of Samos (341-270 BC) and the medieval philosopher and theologian Giordano Bruno (1548-1600 AD) imagined habitable worlds around other stars than our sun (Crowe, 1986). These thinkers were following an ancient philosophical and theological tradition, but their ideas that we are not alone in the universe have had to wait for more than two thousand years for the possibility of observational or experimental evidence.

Our understanding of our place in the Universe changed dramatically in 1995, when Michel Mayor and Didier Queloz of Geneva Observatory announced the discovery of an extra-solar planet around a star similar to our Sun (Mayor and Queloz, 1995). Geoff Marcy and Paul Butler (1995) soon confirmed their discovery, and the science of observational extrasolar planetology was born. The field has exploded in the last dozen years, resulting in a large number of published planetary systems (see http://exoplanet.eu/ for an up-to-date list).

Many of these systems contain one or more gas giant planets very close to their parent star (0.02-0.1AU), and thus do not resemble our Solar System. Although very interesting, they do not directly address the possibility of other worlds like our own. Observational techniques continue to mature, however, and planets with a size and mass similar to the Earth may soon be within reach. "Super-Earths" are several times more massive than our planet and some might have life-supporting atmospheres (Lovis *et al*., 2006; Selsis *et al*., 2007a, Tinetti *et al*., 2007, Kaltenegger *et al*., 2008b). Recent examples of super-Earths are Gl 581c, GL 581d (Udry *et al*., 2007, Selsis *et al*., 2007c) and GJ 436b (Butler *et al*. 2004).

Finding Earth analogues in terms of mass and size will be the focus of many ground and space-based research programmes in the coming decade. Finding evidence of habitability and life represents an even more exciting challenge. Semi-empirical estimates exist of the abundance of terrestrial planets, including the frequency of life and technologically advanced civilizations. Some of these assessments are based on the Drake equation. Unfortunately, they are only educated guesses, not because the equation *per se* is incorrect, but rather because nearly all of its factors are essentially undetermined due to the lack of observational tests. Thus, the basic questions remain open: "Are there planets like our Earth out there?" and "Do any of them harbour life?". In a recent article, G. Bignami (2007) stated: "Finding signatures of life on another (unreachable) planet will be the most important scientific discovery of all times, and a philosophical turning point. It will make clear to everyone that science and exploration eventually pay off, that it is only through the intuition of science (and its associated pigheadedness) that we understand our place among the stars."

To characterize terrestrial exoplanets, we need to detect their light and analyse it by spectroscopy. In two spectral domains, the star to planet contrast ratio is most favourable for an Earth-like planet - in the visible where the planet reflects stellar light, and the thermal infra-red (IR) where the planet emits thermal radiation. To extract the faint signal of a terrestrial planet from the associated large flux from its star the planet must be spatially resolved. Considering this constraint in the thermal IR, an interferometer with collectors carried by formation flying spacecraft with baselines adjustable from 20 m to ~ 200 m has been identified as the best suited instrument. The need to carry out these observations from space results from many factors, including the opacity of our atmosphere in several of the key spectral bands where one would want to observe gases as $H_2O$, $O_3$ and $CO_2$. Building on the pioneering efforts of Bracewell (1978) and Angel (1990), a team of European researchers proposed the *Darwin* concept to ESA in 1993 (Léger *et al*., 1996) and it has been studying it since. NASA has been advancing a similar concept, the Terrestrial Planet Finder Interferometer (TPF-I), since 1996 (Beichman *et al*., 1999). The present *Darwin* concept represents the cumulative effort and synthesis of these studies.

This paper describes the science programme and some of the technological requirements for an ambitious space mission to discover and characterize Earth-like planets and to search for evidence of life on them. The *Darwin* mission will address one of the most fundamental questions: humankind's origin and place in the Universe.

**THE *DARWIN* SCIENCE PROGRAM**

Searching for a phenomenon as subtle as life across parsecs of empty space may look hopeless at first glance, but considerable observational, laboratory, and theoretical effort over the past two decades is leading to the consensus that this is feasible in the near future.

To approach the observational challenge of searching for life we must first address the question, 'what is life?' A living being is a system that contains information and is able to replicate and evolve through random mutation and natural (*Darwin*ian) selection (Brack, 2007). Although this definition appears overly generic (for example, it includes some computer viruses), consideration of possible storage media for life's information leads to a number of specific conclusions.

Macromolecules appear to be an excellent choice for information storage, replication, and evolution in a natural environment. Specifically, carbon chemistry is by far the richest and most flexible chemistry. The need for rapid reaction rates between macromolecules argues for a liquid solution medium. Based on physical and chemical properties as well as abundances in the universe, the most favourable, although not necessarily unique, path for life to take is carbon chemistry in a water solution (Owen, 1980). Our search for signs of life is therefore based on the assumption that extraterrestrial life shares fundamental characteristics of life on Earth, in that it requires liquid water as a solvent and has a carbon-based chemistry (Owen, 1980; Des Marais *et al*., 2002). Such chemistry produces a number of detectable gaseous biological indicators in the planet's atmosphere. We assume that extraterrestrial life is similar to life on Earth in its use of the same input and output gases, that it exists out of thermodynamic equilibrium, and that it has analogs to microorganisms on Earth (Lovelock, 1975, 2000). This assumption is currently necessary because we have no test cases for entirely novel types of life, although any gases out of geochemical equilibrium in a planetary atmosphere might suggest the presence of life.

The logic of the *Darwin* science program follows directly: to search for habitable planets – those where liquid water can exist – and investigate their atmospheres for biosignatures, the gas products known to be produced by the carbon macromolecule chemistry we call life.

*The scientific context*

Since the discovery of a planet orbiting the star 51 Pegasi (Mayor and Queloz, 1995), many planets outside our own Solar System have been discovered. These planets are found in a variety of environments. The majority of these have gas giant masses that cover masses in the range 20 – 3 000 $M_\oplus$. Many of them are either in highly eccentric or very small (0.1 - 0.02 AU) orbits (Udry and Santos, 2007). The latter have surface temperatures up to 2 000 K, and are hence known as "Hot Jupiters".

The existence of Hot Jupiters can be explained by inward migration of planets formed at larger distances from their star, most likely due to tidal interactions with the circumstellar disk. We have also learned that giant planets preferentially form around higher metallicity stars: almost 15% of solar-type stars with metallicity greater than 1/3 that of the Sun possess at least one planet of Saturn mass or larger (Santos *et al*., 2004; Fischer and Valenti, 2005).

Despite substantial efforts, no Earth-mass planet around a normal star has yet been found because of biases towards high mass planets and sensitivity limitations for planet searches from the ground; the lowest mass exoplanets range from 5 to 7 $M_\oplus$.

Planets form within disks of dust and gas orbiting newly born stars. Even though not all aspects are yet understood, growth from micrometer dust grains to planetary embryos through collisions is believed to be the key mechanism leading to the formation of at least terrestrial planets, and possibly the cores of gas giants (Wetherill and Stewart, 1989; Pollack *et al*., 1996).

As these cores grow, they eventually become massive enough to gravitationally bind nebular gas. While this gas accretion proceeds slowly in the early phases, it eventually leads to a runaway buildup of material, once a critical mass has been reached (~10 $M_\oplus$), allowing the formation of a gas giant within the lifetime of the gaseous disk (Pollack *et al.,* 1996). Terrestrial embryos, being closer to the star, have less material available and hence they empty their feeding zone before growing massive. They must then rely on distant gravitational perturbations to induce further collisions. As a result, the growth of terrestrial planets occurs on longer timescales than for the giants.

Extended core-accretion models (Alibert *et al*., 2005) can now be used to compute synthetic planet populations, allowing a statistical comparison with observations (see Fig. 1; Ida and Lin, 2004; Benz *et al.*, 2007). While these models are not specific to terrestrial planets (they are initialized with a seed of 0.6 $M_\oplus$), they demonstrate that if planetary embryos can form, only a small fraction of them will grow fast enough and big enough to eventually become giant planets. Given that we detect gas giants orbiting about 7% of the stars surveyed, *Darwin*'s harvest of terrestrial planets should be very significant.

*Habitability*

The circumstellar Habitable Zone (HZ) (Kasting *et al*., 1993) is defined as the annulus around the star within which starlight is sufficiently intense to maintain liquid water at the surface of the planet. Here we do consider liquid water outside this classical habitable zone, such as on Europa-like moons, as these would not be detected by Darwin. Nevertheless one should note here the potential for liquid water on planetary bodies outside this classical definition of the habitable zone. The inside boundary of the habitable zone is generally set by the location where a runaway greenhouse condition causes dissociation of water and sustains the loss of hydrogen to space and the outer zone is where a maximum greenhouse effect keeps the surface of the planet above the freezing point on the outer edge. In this paper, this inner zone is taken to be 0.84 AU as a conservative estimate, although clouds and water vapour could extent the habitable zone further in. The defined HZ around a certain star implicitly assumes that habitability requires planets of the size and mass similar to the Earth, with large amounts of liquid water on the surfaces, and having $CO_2$ reservoirs with $CO_2$-$H_2O$-$N_2$ atmospheres (Selsis *et al.*, 2007b). The Continuously HZ is the region that remains habitable for durations longer than 1 Gyr. Fig. 2. shows the limits of the HZ as a function of the stellar mass.

Planets inside the HZ are not necessarily habitable. They can be too small, like Mars, to maintain active geology and to limit the gravitational escape of their atmospheres. They can be too massive, like HD69830d, and have accreted a thick $H_2$-He envelope below which water cannot be liquid. However, planet formation models predict abundant Earth-like planets with the right range of masses (0.5 - 8 $M_\oplus$) and water abundances (0.01-10% by mass) (Raymond *et al.,* 2006, 2007a).

To know whether a planet in the HZ is actually inhabited, we need to search for biosignatures, features that are specific to biological activities and can be detected by remote sensing. An example is $O_2$-producing photosynthesis. Owen (1980) suggested searching for $O_2$ as a tracer of life. In the particular case of Earth, most $O_2$ is produced by the biosphere. Less than 1 ppm comes from abiotic processes (Walker, 1977). Cyanobacteria, algae and plants are responsible for this production by using solar photons to drive the electron transport chain using water as an electron donor resulting in the production of oxygen (and generating organic molecules from $CO_2$ in associated dark reactions). This metabolism is called oxygenic photosynthesis. The reverse reaction, using $O_2$ to oxidize the organics produced by photosynthesis, can occur abiotically when organics are exposed to free oxygen, or biologically by respiration consuming organics. Because of this balance, the net release of $O_2$ in the atmosphere is due to the burial of organics in sediments. Each reduced carbon buried frees an $O_2$ molecule in the atmosphere. This net release rate is also balanced by weathering of fossilized carbon when exposed to the surface. The oxidation of reduced volcanic gasses such as $H_2$, $H_2S$ also accounts for a significant fraction of the oxygen losses. The atmospheric oxygen is recycled through respiration and photosynthesis in less than 10,000 yrs. In the case of a total extinction of Earth's biosphere, the atmospheric $O_2$ would disappear in a few million years.

Chemolithotrophic life, thriving in the interior of the planet without using stellar light, can still exist outside (as well as inside) the HZ. The associated metabolisms – at least the ones we know on Earth – do not produce oxygen. By comparison, the energy and electron donors for photosynthesis, sunlight and water respectively, are widely distributed, yield larger biological productivity and can modify a whole planetary environment in a detectable way (Léger *et al*., 1993; Wolstencroft and Raven, 2002; Raven and Wolstencroft, 2004; Kiang *et al*., 2007a,b; Cockell, 2008). Light sources to sustain photosynthesis are likely to be widely available in different planetary systems (Raven and Cockell, 2006), albeit they also are associated with different and potentially more hostile ultraviolet (UV) radiation regimens (Kasting *et al*., 1997; Cockell, 1999; Segura *et al.*, 2003). For these reasons, searching for oxygenic photosynthesis is a restricted way to search for life on planets within the HZ of their stars, but it is based on empirical free energy considerations concerning the likely impact on atmospheric composition. Most importantly, it is a search that can be done by remote sensing, looking for spectroscopic signatures of $O_2$ or its tracers e.g. $O_3$.

*Search for biosignatures*

The range of characteristics of planets found in the HZ of their star is likely to greatly exceed our experience with the planets and satellites in our Solar System. In order to study the habitability of the planets detected by *Darwin*, and to ascertain the biological origin of the measured atmospheric composition, we need a comprehensive picture of the observed planet and its atmosphere.

In addition to providing a more favourable planet-star contrast and potential biosignatures, the mid-IR (MIR) wavelength domain provides crucial chemical, physical and climatic diagnostics, even at moderate spectral resolution. For example, the infrared light curve, i.e. the variation of the integrated thermal emission with location on the orbit, reveals whether the detected planet possesses a dense atmosphere, suitable for life, which strongly reduces the day/night temperature difference.

A low-resolution spectrum spanning the 6-20 μm region allows us to measure the effective temperature $T_{eff}$ of the planet, and thus derive its radius $R_{pl}$ and albedo (see section 'Mission performance'). Low-resolution mid-IR observations will also reveal the existence of greenhouse gases, including $CO_2$ and $H_2O$.

Within the HZ, the partial pressure of $CO_2$ and $H_2O$ at the surface of an Earth-analogue is a function of the distance from the star. Water vapour is a major constituent of the atmosphere for planets between 0.84 AU and 0.95 AU. Fig. 3. shows the estimated changes of the $H_2O$, $O_3$ and $CO_2$ features in the spectra of an Earth-like planet as a function of its location in the HZ. Carbon dioxide is a tracer for the inner region of the HZ and becomes an abundant gas further out (Kaltenegger and Selsis, 2007, Selsis *et al*., 2007b).

Planets such as Venus, closer to their star than the HZ, can lose their water reservoir and accumulate a thick $CO_2$ atmosphere. Such planets can be identified by the absence of water and by the $CO_2$ absorption bands between 9 and 11 μm, which only show up in the spectrum at high $CO_2$ mixing ratios. These conditions would suggest an uninhabitable surface.

*Darwin* will test the theory of habitability indicators versus orbital distance by correlating the planets' spectral signature with orbital distance and comparing the results with grids of theoretical spectra, such as those shown in Fig. 3.

Fig. 4. shows that the mid-IR spectrum of Earth displays the 9.6 μm $O_3$ band, the 15 μm $CO_2$ band, the 6.3 μm $H_2O$ band and the $H_2O$ rotational band that extends beyond 12 μm. The Earth's spectrum is clearly distinct from that of Mars and Venus, which display the $CO_2$ feature only. Fig 5. illustrates the physical basis behind the spectra shown in Fig 4.

The combined appearance of the $O_3$, $H_2O$, and $CO_2$ absorption bands is the best-studied signature of biological activity (Léger *et al*., 1993; Schindler and Kasting, 2000; Selsis *et al*., 2002; Des Marais *et al*., 2002; Segura *et al*., 2007). Despite variations in line shape and depth, atmospheric models demonstrate that these bands could be readily detected with a spectral resolution of 10–25 in Earth analogues covering a broad range of ages and stellar hosts (Selsis, 2000; Segura *et al*., 2003; Kaltenegger *et al*., 2007).

The ozone absorption band is observable for $O_2$ concentrations higher than 0.1% of the present terrestrial atmospheric level (Segura *et al*., 2003). The Earth's spectrum has displayed this feature for the past 2.5 Gyr.

Other spectral features of potential biological interest include methane ($CH_4$ at 7.4 µm), and species released as a consequence of biological fixation of nitrogen, such as ammonia ($NH_3$ at 6 and 9-11 µm) and nitrous oxide ($N_2O$ at 7.8, 8.5 and 17 µm). The presence of these compounds would be difficult to explain in the absence of biological processes. Methane and ammonia commonly appear in cold hydrogen-rich atmospheres, but they are not expected as abiotic constituents of Earth-size planetary atmospheres at habitable orbital distances.

Methane, ammonia and nitrous oxide do not produce measurable spectral signatures at low resolution for an exact Earth analogue. Nevertheless, they may reach observable concentrations in the atmosphere of exoplanets, due either to differences in the biosphere and the planetary environment, or because the planet is observed at a different evolutionary phase, as illustrated in Fig. 6. Methane, for instance, was most likely maintained at observable concentrations for more than 2.7 Gyrs of Earth's history from about 3.5 until 0.8 Gyrs ago (Pavlov *et al*., 2003). During the 1.5 Gyrs following the rise of oxygen (2.4 Gyrs ago), the spectrum of the Earth featured deep methane absorption simultaneously with ozone. The detection of reduced species, such as $CH_4$ or $NH_3$, together with $O_3$, is another robust indicator of biological activity (Lovelock, 1975, 2000; Sagan *et al*., 1993).

The presence of $H_2O$, together with reduced species such as $CH_4$ or $NH_3$, would also be indicative of a possible biological origin. Although a purely abiotic scenario could produce this mix of gases, such a planet would represent an important astrobiological target for future study. The presence of nitrous oxide ($N_2O$) and, more generally, any composition that cannot be reproduced by a self-consistent abiotic atmosphere model would merit follow-up.

Finally, if biology is involved in the geochemical cycles controlling atmospheric composition, as on Earth, greenhouse gases will probably be affected and sustained at a level compatible with a habitable climate. Whatever the nature of these greenhouse gases, *Darwin* will be able to see their effect by analyzing the planet's thermal emission. This is a powerful way to give the instrument the ability to characterize unexpected worlds.

**COMPARATIVE (EXO) PLANETOLOGY**

Over the decade since the discovery of 51 Peg, we have grown to understand that planetary systems can be much more diverse than originally expected (Udry and Santos, 2007). It is also clear that the current group of extrasolar planets, although diverse, is incomplete: as observational techniques have improved, we have pushed the lower limit of the detected masses closer and closer to the terrestrial range. In the coming decade, this trend will continue, and our understanding of the diversity of lower mass planets will be critical to the understanding of the formation of terrestrial planets in general, and of the Earth in particular.

Growing the sample of terrestrial planets from the three in our solar system to a statistically significant sample will represent a quantum leap in knowledge. And, just as 51 Peg created the discipline of observational extrasolar planetology, this effort will encompass a new type of science: comparative exoplanetology for both the giant and terrestrial planets. It will allow for the first time a comparison of the orbital, physical and chemical characteristics of full planetary systems with our solar system and model predictions (Selsis *et al*., 2007b).

Finally, this sample will also allow help answer one of the key questions related to *Darwin* science: How frequently are planets, which are located in or near the HZ, truly habitable?

*Darwin* can determine the radius, but not the mass, of planets. Ground-based radial velocity measurements can provide this information, however. The estimated error in mass determination is a function of planet mass, stellar type, visual magnitude, etc. Achieving adequate mass accuracy will be possible with instruments such as HARPS on 8-meter class telescopes for a fraction of the discovered planets (Li *et al*., 2008). Large planets, e.g. 2 – 5 $M_\oplus$, around small stars that are nearby and therefore bright – mainly M and K stars - are the best candidates.

The origin and evolution of liquid water on the Earth is an ideal example of the type of puzzle that comparative exoplanetology will address. Our planet orbits in the Habitable Zone of our star, but at least some of the water on Earth must have been delivered by primordial icy planetesimal and/or water rich chondritic meteorites.

Did the early Earth capture these objects when they wandered into the inner solar system, or did our planet itself form further out and migrate inward? The answer is not clear at this point. What is clear is that habitability cannot just be reduced to a question of present-day location. The origin and fate of the water reservoir within the proto-planetary nebula is equally important.

By necessity, we have until now addressed this question using the very restricted sample of terrestrial planets in our own solar system: Venus, Earth and Mars. What have we learned? The in situ exploration of Mars and Venus taught us that all three planets probably evolved from relatively similar initial atmospheric conditions, most probably including a primordial liquid water reservoir. In all three cases, a thick $CO_2$ atmosphere and its associated greenhouse effect raised the surface temperature above the classical radiative equilibrium level associated with their distance to the Sun. This atmospheric greenhouse effect was critical for habitability on Earth at a time when the young Sun was approximately 30% fainter than it is today (Newman and Rood, 1977; Gough, 1981)

At some point in the past, the evolutionary paths of Venus, Earth, and Mars began to diverge. For Venus, the combination of the greenhouse effect and a progressively hotter Sun led to the vaporization of all liquid water into the atmosphere, assuming a similar water reservoir as Earth. After upward diffusion, $H_2O$ was dissociated by UV radiation, causing the loss of hydrogen to space. Venus is today a hot, dry, and uninhabitable planet.

In contrast, Mars apparently experienced a 500 million year episode with a warmer, wetter climate, before atmospheric loss and a steady decrease in surface temperature trapped the remaining water reservoir in the polar ice caps and subsurface permafrost. Thus, Mars retained a fraction of its water reservoir.

Earth apparently followed an intermediate and complex evolutionary path, which maintained its habitability for much of the past 4.6 Gyrs. Early on, a thick $CO_2$ atmosphere, possibly combined with other greenhouse gases, compensated for the young Sun's low luminosity, and, as the Sun brightened, atmospheric $CO_2$ was progressively segregated into carbonate rocks by the combined action of the water cycle, erosion, sedimentation of carbonate deposits on the ocean floors, and partial recycling via plate tectonics. This feedback cycle, which appears unique in the Solar System, accounts for the preservation of Earth's oceans and habitability throughout most of its history (Kasting and Catling, 2003).

Although the general definition of the HZ can be applied to all stellar spectral types one can expect that the evolution of the atmospheres of terrestrial planets within the HZs of lower mass M and K-type stars is different. These planets have closer orbital locations and experience denser stellar plasma environments (winds, Coronal Mass Ejections) (Khodachenko *et al*., 2007; Lammer *et al*., 2007; Scalo *et al*., 2007). These stars have longer active strong X-ray and extreme UV periods compared to solar-like stars. In addition, planets in orbital locations in the HZ of low mass stars can be partially or totally tidal-locked, which could result in different climate conditions, and weaker magnetic dynamos important for protecting the atmosphere against erosion by the stellar plasma flow. These differences compared to G-star HZ planets raise questions regarding atmospheric escape, plate tectonics, magnetic dynamo generation and the possibility of complex biospheres (Grießmeier *et al*., 2005).

Terrestrial planets with no analogue in the Solar System may exist. For example, the recently proposed Ocean-Planets, which consist of 50% silicates and 50% water in mass (Léger *et al*., 2004; Selsis *et al*., 2007a), could form further out than the distance from the star where water vapour condenses (~ 4 AU around a G dwarf) and migrate to the HZ, or closer. Such objects would be a new class of planets, the terrestrial analogues of hot Jupiters and Neptunes. If these planets happen to exist, *Darwin* will be able to characterize them in detail, e.g. determining their atmospheric compositions.

With *Darwin*, the sample of terrestrial planets will be extended to our galactic neighbourhood, allowing us to study the relationship between spectral characteristics and three families of parameters:

- Stellar characteristics, including spectral type, metallicity, and if possible, age; our Solar System illustrates the importance of understanding the co-evolution of each candidate habitable planet and its star.
- Planetary system characteristics, particularly the distribution and the orbital characteristics of terrestrial and gas giant planets.
- The atmospheric composition of planets in the HZ. Here again, the solar system sample points to the importance of ascertaining the relative abundance of the main volatile species: $CO_2$, $CH_4$, $H_2O$, $O_3$, $NH_3$, etc.

The strategy for comparative exoplanetology will be as follows: First, a comparison of stellar characteristics with the nature of the planetary system will capture the diversity of planetary systems. Then, *Darwin*'s spectroscopic data will reveal the range of atmospheric compositions in the Habitable Zone, a range that will be related to the initial chemical conditions in the proto-planetary nebula and, if stellar ages are available, to the effects of atmospheric evolution.

Correlating the general characteristics of the planetary system with the atmospheres of the individual planets will illuminate the interplay between gas giants and terrestrial bodies and the role of migration. For example, recent numerical simulations predict that the scattering effect of giant planets on the population of planetesimals plays a key role in the collisional growth of terrestrial planets, their chemical composition and the build-up of their initial water reservoir (Raymond *et al*., 2006b).

Thus, *Darwin* will allow us to address the question of habitability from the complementary perspectives of the location of Earth-like planets with respect to their HZ, and of the origin, diversity and evolution of their water reservoirs.

**HIGH ANGULAR RESOLUTION ASTRONOMY WITH *DARWIN***

*Darwin*'s long interferometric baselines and large collecting area make it a powerful instrument for general astrophysics. The mission combines the sensitivity of JWST (James Webb Space Telescope, Gardner et al. 2006) with the angular resolution of VLTI (Very Large Telescope Interferometer, Schöller *et al*., 2006) in an instrument unencumbered by atmospheric opacity and thermal background.

The baseline instrumentation in *Darwin* will be able to observe general astrophysics targets whenever there is a bright point source, e.g. a non-resolved star in the field of view. This source is necessary to cophase the input pupils. Some science programs will profit from just a few visibility measurements, while others will require numerous observations and complete aperture synthesis image reconstruction.

Taking full advantage of interferometric imaging with *Darwin* – that is observing any source on the sky – will require specialized and potentially costly add-on instrumentation to allow the cophasing of the array. General astrophysics is not the primary science mission. However, it should be remembered that general astrophysical studies can in themselves improve our understanding of the astrobiological environment for life and the nature of stars around which life might exist. Therefore, for completeness, it is useful here to briefly summarise some of these other astrophysical goals.

**Star Formation**

Stars are the fundamental building blocks of the baryonic universe. They provide the stable environment needed for the formation of planetary systems and thus, for the evolution of life. *Darwin* will impact our understanding of star formation in fundamental ways, for instance the 'Jet-Disk Connection'. Forming stars launch powerful jets and bipolar outflows along the circumstellar disk rotation axis (Shu *et al*., 1987; Reipurth and Bally, 2001). The mission could reveal the nature of the driving mechanism by spatially resolving the jet-launching region. Are jets formed by ordinary stellar winds, the magnetic X-points where stellar magnetospheres interact with the circumstellar disk, or are they launched by magnetic fields entrained or dynamo-amplified in the disk itself?

**Planet Formation**

Theory predicts that planets form in circumstellar disks over a period of $10^6$ to $10^8$ years (Pollack *et al*., 1996; Boss, 1997). *Darwin* can provide detailed information about planetary systems at various stages of their evolution, revealing the origin of planetary systems such as our own, and thus helping to place our solar system into context. The mission will be unique in being able to spatially resolve structures below 1 AU in nearby star forming regions, allowing us to witness directly the formation of terrestrial planets in the thermal IR. Additional planet formation science includes:

- *Disk formation and evolution.* *Darwin* will place constraints on the overall disk structure. The mission measurements will directly constrain grain growth, settling, and mixing processes in the planet-forming region (see for example, Hollenbach et al. 2000; Rieke *et al.*, 2005).

- *Disk Gaps within the Inner Few AU.* The mid-IR spectral energy distribution of protoplanetary and debris disks points to the existence of gaps. *Darwin* will determine if this clearing is due to the influence of already-formed giant planets or if they are the result of viscous evolution, photo-evaporation, and dust grain growth (see, for example, Dominik and Decin, 2003; Throop and Bally, 2005).

**Formation, Evolution, and Growth of Massive Black Holes**

How do black holes (BH) form in galaxies? Do they form first, and trigger the birth of galaxies around them, or do galaxies form first and stimulate the formation of BHs? How do BHs grow? Do they grow via mergers as galaxies collide? Or do they accumulate their mass by hydrodynamic accretion from surrounding gas and stars in a single galaxy? *Darwin* could probe the immediate environments of very different BHs, ranging from very massive BHs in different types of active galactic nuclei (AGN), to the massive black hole at the centre of our own Milky Way, down to BHs associated with stellar remnants.

*Darwin* will be able to make exquisite maps of the distribution of silicate dust, ices, and polycyclic aromatic hydrocarbons (PAHs) in weak AGN such as NGC 1068 out to a redshift of $z=1-2$. Brighter AGN can be mapped to a redshift of $z=10$, if they exist. For the first time, we will be able to measure how the composition, heating, and dynamics of the dust disks change with redshift (Dwek, 1998; Edmunds, 2001). This will provide a clear picture of when and how these tori and their associated massive BHs grow during the epoch of galaxy formation (Granato *et al.*, 1997).

*The Galactic centre:* The centre of our Galaxy contains the nearest massive black hole (3.6 x $10^6$ M$_\odot$)(e.g., Ghez *et al*., 2005) , a uniquely dense star cluster containing more that $10^7$ stars pc$^{-3}$, and a remarkable group of high-mass stars (e.g., IRc7, Yusef-Zadeh and Morris, 1991). *Darwin* will be able to trace the distribution of lower mass stars and probe the distribution of dust and plasma in the immediate vicinity of the central BH.

**Galaxy Formation & Evolution**

Galaxy evolution is a complicated process, in which gravity, hydrodynamics, and radiative heating and cooling all play a fundamental role (Dwek, 1998; Edmunds, 2001). Measurements of the detailed spatial structure of very distant galaxies will place essential constraints on galaxy formation models. Unlike JWST, *Darwin* will be able to resolve individual OB associations, massive star clusters, and their associated giant HII regions. By carefully selecting targets of a specific type, we can trace the evolution of galaxy structure as a function of redshift and environment (Franx *et al*., 2003; Labbé *et al*., 2005). The evolution of metallicity with cosmic age and redshift can be mapped using various diagnostics, including molecular tracers, ices, PAH bands, and noble gas lines that are in the (6 – 20 µm) band (e.g. Moorwood *et al*., 1996; Soifer *et al*., 2004). Fig. 7 shows an example of the mapping power of the mission.

**The First Generation of Stars**

The first stars formed in the early universe are thought to be quite different from the stars present today. The absence of metals reduced the opacity, allowing the first generation of stars to accumulate more gas and hence be considerably more massive (100 to 1000 M$_\odot$) and hotter than their modern counterparts (Brom *et al*., 1999; Tumlinson and Shull, 2000). The first stars must have had a dramatic impact on their environment, creating giant HII regions whose red-shifted hydrogen and helium emission lines should be readily observable by *Darwin*. While JWST is expected to make the first detections of galaxies containing these "Population III" stars (e.g., Gardner *et al.*, 2006), *Darwin* will be capable of resolving scales of order 10 to 100 pc at all redshifts, providing the hundred-fold gain needed to resolve these primordial HII regions. *Darwin* will also be able to test the current paradigm for the formation of the first stars. Are they truly isolated, single objects that have inhibited the formation of other stars in their vicinity, or are they surrounded by young clusters of stars?

**Other Important Science**

The mission general astrophysics program could include a number of additional key science programmes:

- *Our home planetary system:* *Darwin* will be able to easily measure the diameters, and properties of Kuiper Belt Objects, moons, asteroids, and cometary nuclei. Low-resolution spectro-photometry will constrain the natures of their surfaces, atmospheres, and environments.

- *AGB stars:* *Darwin* can provide detailed maps of the distribution of dust and gas within the envelopes of oxygen-rich (M-type) and carbon-rich (C-type) AGB stars, in environments as extreme as the Galactic centre.

- *Supernovae:* *Darwin* could image the formation and evolution of dust, atoms and ions in supernova ejecta, and trace the structure of the circumstellar environment into which the blast is propagating.

- *Dark matter & dark energy:* *Darwin* studies of gravitational lensing by galaxy clusters, AGN, and ordinary galaxies could place unprecedented constraints on the structure of dark matter haloes at sub-kpc scales.

**SYNERGIES WITH OTHER DISCIPLINES**

The primary *Darwin* science objective is inherently multi-disciplinary in character, uniting astronomy with chemistry, geology, physics and branches of biology, including microbiology. Often referred to as astrobiology, this interdisciplinary field also includes molecular biology, celestial mechanics and planetary science, including the physics and chemistry of planetary atmospheres and the characterization of exoplanetary surfaces. Climatologists and ecologists will have the opportunity to study global climate influences within the context of a statistically large number of other terrestrial planets. Of particular interest will be an understanding of the diversity of Venus-like planets undergoing runaway greenhouse changes. On the technological front, the mission will drive development in such widely differing fields as material sciences, optical design, and spacecraft Formation Flying.

**THE *DARWIN* MISSION PROFILE**

*Baseline Mission Scope*

The *Darwin* mission consists of two phases: the search for and spectral characterization of habitable planets, whose relative duration can be adjusted to optimize scientific return. During the search phase (nominally 2 yrs), the mission will examine nearby stars for evidence of terrestrial planets. An identified planet should be observed at least 3 times during the mission in order to characterize its orbit. The number of stars that can be searched depends on the level of zodiacal light in the system and the diameter of the collector telescopes. As a baseline, we estimate this number under the assumption of a mean exozodiacal density 3 times that in the Solar System and collecting diameters of 2 m. Over 200 stars can be screened under these conditions. The mission focuses on Solar type stars which are long-lived, i.e., F, G, K and some M spectral types (Kaltenegger *et al*., 2008a).

The number of expected planetary detections depends upon the mean number of terrestrial planets in the HZ per star, $\eta_\oplus$. Our present understanding of terrestrial planet formation and our Solar System, where there are 2 such planets (Earth & Mars) and one close to the HZ (Venus), point to a fairly high abundance of terrestrial planets. We assume hereafter that $\eta_\oplus = 1$ for simulations. The COROT mission should reveal the abundance of small hot planets, and *Kepler* will evaluate $\eta_{Earth}$ as well as the size distribution of these objects several years before *Darwin* flies. These inputs will allow refinement of *Darwin*'s observing strategy well in advance of launch.

During the characterization phase of the mission (nominally 3 yrs), *Darwin* will acquire spectra of each detected planet at a nominal resolution of 25 and with sufficient signal-to-noise (~10) to measure the equivalent widths of $CO_2$, $H_2O$, and $O_3$ with a precision of 20% if they are in abundances similar to those in the present-day Earth atmosphere.

Spectroscopy is more time consuming than detection. Assuming $\eta_\oplus = 1$, *Darwin* can perform spectroscopy of $CO_2$ and $O_3$ on about 50 earth-analogues and of $H_2O$ on about 25 during the nominal 3-year characterization phase.

The general astrophysics program, if adopted, will comprise 10% to 20% of the mission time. The primary science segment would then be reduced accordingly, with limited impact on its outcome.

*Extended Mission Scope*

An extension of the mission to 10 years will depend on the results gathered during the first 5 years. Such an extension could be valuable to observe more M stars; only 10% of the baseline time is attributed to them currently. As they are the most stable stars, the chances for habitable planets are good (Tarter *et al*., 2007). An extension would also allow for more observation of large planets around a significantly larger sample of stars. A major reason to extend the mission will be to make additional measurements on the most interesting targets already studied and to expand the range of masses and environments explored.

*Darwin* Target Catalogue.

The *Darwin* target star catalogue (Kaltenegger, 2005; Kaltenegger *et al*., 2008a) is generated from the Hipparcos catalogue by examining the distance (< 30 pc), brightness, spectral type (F, G , K, M main sequence stars), and multiplicity (no companion). The catalogue has considered different interferometer architectures, since they have different sky access. The Emma design can observe 99% of the sky (see below). The corresponding star catalogue contains 1256 single target stars within 30 pc, 414 of the 1256 target stars are M stars, 515 are K stars, 218 are G stars and 109 are F stars. Of the 1256 stars, 36 are known to host exo-planets. Fig. 8 shows some features of these stars.

## *DARWIN* MISSION DESIGN

*The Darwin Concept and Its Evolution*

In order to disentangle the faint emission of an Earth-like planet from the overwhelming flux of its host star, the planetary system needs to be spatially resolved. This, in turn, requires a monolithic telescope up to 100 m in diameter when operating at mid-IR wavelengths since the angular size of the Habitable Zones around *Darwin* target stars ranges between 10 and 100 milliarcseconds (mas, see Fig. 8). A telescope of this size is presently not feasible, particularly since the observatory must be space-borne and cooled to provide continuous coverage and sensitivity between 6 and 20 µm.

As a result, interferometry has been identified as the best-suited technique to achieve mid-IR spectroscopy of Earth-like planets around nearby stars. In his pioneering paper, Bracewell (1978) suggested that applying a $\pi$ phase shift between the light collected by two telescopes could be used to cancel out the on-axis star, while allowing the signal from an off-axis planet to pass through (Fig. 9). This technique, referred to as *nulling interferometry*, has been at the heart of the *Darwin* concept since its origin (Léger *et al*., 1996) and many improvements have been studied since that date.

In addition to the planetary flux, a number of spurious sources contribute to the signal at the destructive output of the Bracewell interferometer (Mennesson *et al*., 2005; Absil *et al*., 2006):

- Residual star light, referred to as *stellar leakage*, caused by the finite size of the stellar photosphere (geometric leakage) and by the imperfect control of the interferometer (instrumental leakage);
- The *local zodiacal background*, produced by the disk of warm dust particles that surround our Sun and radiate at infrared wavelengths;
- The *exozodiacal light*, arising from the dust disk around the target star;

- The *instrumental background* produced by thermal emission within the instrument.

Bracewell's original suggestion of rotating the array of telescopes can help disentangle the various contributions. The planet signal would then be temporally modulated by alternatively crossing high and low transmission regions, while the stellar signal and the background emission remain constant (except for the exozodiacal emission). Unfortunately, this level of modulation is not sufficient to achieve *Darwin's* goals, prompting a series of improvements to the strategy, including:

- Breaking the symmetry of the array to cancel all centro-symmetric sources, including the geometric stellar leakage, the local and exozodiacal emissions;

- Performing faster modulation of the planet signal via phase modulation between the outputs of sub-interferometers.

Merging of these two ideas has led to the concept of *phase chopping* (Woolf & Angel, 1997; Mennesson *et al.*, 2005) which is now regarded as a mandatory feature in space-based nulling interferometry. The simplest implementation of phase chopping is illustrated in Fig. 10: the outputs of two Bracewell interferometers are combined with opposite phase shifts ($\pm\pi/2$) to produce two "chopped states," which are mirrored with respect to the optical axis. Taking the difference of the photon rates obtained in the two chopped states gives the chopped response of the array, represented by the modulation map. This chopping process removes all centro-symmetric sources, including the geometric stellar leakage and the exozodiacal emission.

Because the modulation efficiency varies across the field-of-view, the planet can only be localised and characterised through an additional level of modulation, provided by array rotation with a typical period of one day. The collected data, consisting in time series of detected photo-electrons at the destructive and constructive outputs of the interferometer, must be inverted to obtain the fluxes and locations of any planets that are present. The most common approach is correlation mapping, which is closely related to the Fourier transform used for standard image synthesis. The result is a correlation map, which represents the point spread function (PSF) of the array (Fig 10).

This process is repeated across the waveband, and the maps are co-added to obtain the net correlation map. The broad range of wavelengths planned for *Darwin* greatly extends the spatial frequency coverage of the array, suppressing the side lobes of the PSF.

A dozen array configurations using phase chopping have been proposed and studied during the past decade (Mennesson and Mariotti, 1997; Karlsson and Mennesson, 2000; Absil *et al*., 2003; Karlsson *et al*., 2004; Kaltenegger and Karlsson, 2004; Lay, 2005; Mennesson *et al*., 2005). In 2004, the ESA and NASA agreed on common figures of merit to evaluate their performance. The most important criteria are the modulation efficiency of the beam combination scheme, the structure of the PSF and its associated ability to handle multiple planets, the overall complexity of beam routing and combination, and finally, the number of stars that can be surveyed during the mission lifetime (Lay, 2005).

*Mission Architecture*

The desire for maximum mission efficiency, technical simplicity, and the ability to detect multiple planets around as many stars as possible has guided the selection of mission architecture. Additional top-level requirements include:

- Placement at L2 for passive cooling and low ambient forces;
- Launch with a single Ariane 5 rocket or two Soyuz-ST/Fregat vehicles;

- The ability to search a statistically meaningful sample of nearby solar-type stars (~200) for the presence of habitable planets, assuming an exozodiacal background up to 10 zodi[1];

- The ability to detect and measure terrestrial atmosphere biosignatures for a significant fraction of the planets found during the search phase (at least 20);

- Time allocation during search phase as follows: G stars 50%, K stars 30%, F and M stars 10% each;

- Two observing modes: nulling for extrasolar planet detection and spectroscopy, and constructive imaging for general astrophysics.

The effort to turn these requirements into a workable mission culminated in 2005-2006 with two parallel assessment studies of the *Darwin* mission. Two array architectures have been thoroughly investigated during these studies: the 4-telescope X-array (Lay and Dubovitsky, 2004) and the 3-telescope TTN (Karlsson *et al.*, 2004). These studies included the launch requirements, payload spacecraft, and the ground segment during which the actual mission science would be executed. Almost simultaneously, NASA/JPL initiated a similar study (Martin *et al.*, 2007) in the context of the Terrestrial Planet Finder Interferometer (TPF-I).

These efforts on both sides of the Atlantic have resulted in a convergence and consensus on mission architecture. The baseline for *Darwin* is a *non-coplanar,* or *Emma*[2]-type X-array, with four Collector Spacecraft (CS) and a single Beam Combiner Spacecraft (BCS). This process also identified a back-up option, in case unforeseen technical obstacles appear: a planar X-array.

---

[1] A "zodi" is defined as the density of our local zodiacal dust disk and acts as a scaling factor for the integrated brightness of exozodiacal dust disks.
[2] Emma was the wife of Charles Darwin.

*The Emma X-Array Architecture*

Fig. 11. shows the non-coplanar Emma X-array. Four simple collector spacecrafts fly in a rectangular formation and feed light to the beam combiner spacecraft located approximately 1200 m above the array. This arrangement allows baselines between 7 and 168 m for nulling measurements and up to 500 m for the general astrophysics program.

The X-array configuration separates the nulling and imaging functions (Fig 10), thus allowing independent optimal tuning of the shorter dimension of the array for starlight suppression and that of the longer dimension for resolving the planet (Lay and Dubovitsky, 2004; Lay, 2005). Most other configurations are partially degenerate for these functions. The X-array also lends itself naturally to techniques for removing instability noise, a key limit to the sensitivity of *Darwin* (Chazelas, 2006; Lay, 2004; 2006; Lane *et al*., 2006).

The assessment studies settled on an imaging to nulling baseline ratio of 3:1, based on scientific and instrument design constraints. A somewhat larger ratio of 6:1 may improve performance by simplifying noise reduction in the post-processing of science images.

Each of the Collector Spacecraft (CS) contains a spherical mirror and no additional science-path optics (some components may be needed for configuration control). The four CS fly in formation to synthesize part of a larger paraboloid—the *Emma* configuration is a single, sparsely filled aperture. Flexing of the CS primary mirrors or deformable optics within the beam combiner spacecraft will conform the individual spheres to the larger paraboloid.

The Beam Combiner Spacecraft (BCS) flies near the focal point of this synthesized paraboloid. Beam combination takes place on a series of optical benches arranged within the BCS envelope. The necessary optical processing includes:

- Transfer optics and BCS/CS metrology;
- Correction and modulation, including optical delay lines, tip-tilt, deformable mirrors;
- Mirrors, wavefront sensors, and beam switching;
- Spectral separation to feed the science photons into 2 separate channels;
- Phase shifting, beam mixing;
- Recombination, spectral dispersion and detection.

The collector and beam combiner spacecrafts use sunshades for passive cooling to < 50 K. An additional refrigerator within the BCS cools the detector assembly to below 10 K.

Due to the configuration of the array and the need for solar avoidance, the instantaneous sky access is limited to an annulus with inner and outer half-angles of 46° and 83° centred on the anti-sun vector. This annulus transits the entire ecliptic circle during one year, hence giving access to almost the entire sky.

For launch, the collector and beam-combiner spacecraft are stacked within the fairing of an Ariane 5 ECA vehicle. Total mass is significantly less than 6.6 tons, the launcher capability. Table 1 lists key parameters of the *Darwin Emma* X-array. These values represent the results of the various assessment and system level studies conducted by ESA and NASA.

**TABLE 1. KEY *DARWIN* PARAMETERS**

| Item | Value or Comment |
|------|------------------|
| Collector Spacecraft (CS) | 4 free-flyers, passively cooled to <50K |
| CS Optics | Lightweight spherical mirrors, diameter *ca.* 2.0 m, no deployables |
| CS Array Configuration | X-array with aspect ratio 3:1 – 6:1 (to be optimized) |
| Available Baselines | 7 m to 168 m Nulling (20 m to 500 m Imaging option) |
| Beam Combiner (BCS) | 1 free flying spacecraft, passively cooled to <50K |
| Beam Combiner Optics | Transfer, modulation, beam-mixing, recombination, spectroscopy |
| Detection | Mid-IR detector ca. 500 x 8 pixels for nulling, (300 x 300 for imaging option), cooled to < 10 K |
| Detector Cooling | Low vibration refrigerator, e.g. absorption coolers (pulse tube coolers are also possible, e.g., from JWST) |
| Telemetry | Require ca. 1 GBit /s, direct downlink from BCS |
| Operating Wavelength | 6-20 μm. Includes $H_2O$, $O_3$, $CH_4$, $CO_2$ signatures |
| Field of View | Typically 1 arcsec at 10 μm |
| Null Depth | $10^{-5}$, stable over ~ 5 days |
| Angular Resolution | 5 milliarcsec at 10 μm for a 500 m baseline, scales inversely |
| Spectral Resolution | 25 (possibly 300) for exo-planets; 300 for general astrophysics |
| Field of Regard | Annular region between 46° and 83° from anti-sun direction at a date, 99% over one year |
| Target Stars | F, G, K, M, (at least 200) |
| Mission Duration | 5 yrs baseline, extendable to 10 yrs |
| Mission Profile | Nominal 2 yrs detection, 3 yrs spectroscopy, flexible |
| Orbit | L2 halo orbit |
| Formation Flying | Radio Frequency and Laser controlled |
| Station Keeping | Field Effect Electric Propulsion (FEEP) or cold gas |
| Launch Vehicle | Single Ariane 5 ECA or 2 Soyuz-ST / Fregat |

**MISSION PERFORMANCE**

*Detecting Earths*

*Darwin*'s instruments will encounter a number of extraneous signals. The planetary flux must be extracted and analysed in the presence of these other components. The discrimination is performed by maximizing starlight rejection, and by appropriate modulation that produces a zero mean value for the different background sources. Image reconstruction algorithms are then used to retrieve the actual flux and location of the planets, as illustrated in Fig. 12 (e.g. Draper *et al*., 2006; Marsh et al., 2006; Thiébaut and Mugnier, 2006; Mugnier *et al*., 2006). Even though modulation cannot eliminate the quantum noise (sometimes referred to as photon noise) associated with these sources, nor the instability noise related to imperfect instrumental control, these issues have been addressed by Lay (2006) and Lane *et al*. (2006).

*Search Strategy and Performance*

Performance simulations have been conducted for each star in the target catalogue, using both the *Darwin*Sim software developed at ESA (den Hartog, private communication) and the TPF-I star count model developed at NASA (Lay et al., 2007), to assess the required integration time to reach the required signal to noise ratio (SNR) for detection and spectroscopy. These requirements are an SNR of 5 on the whole band for imaging in nulling mode, and a SNR of 10 from 7.2 to 20 µm for $H_2O$, $CO_2$ and $O_3$ spectroscopy, using a spectral resolution $\lambda/\Delta\lambda \geq 20$. Under the assumption that the exozodiacal emission is symmetric around the target star, the chopping process will suppress it, and the exozodi will only contribute to the photon noise. The simulations presented below assume an exozodiacal density of 3 zodis.[3]

Using an Emma X-array (6:1 configuration) with 2-m diameter telescopes and assuming an optical throughput of 10% for the interferometer, we estimate that about 200 stars distributed among the four selected spectral types can be screened during the nominal 2-yrs survey

---

[3] In practice, exozodiacal densities below 10 times our local zodiacal cloud barely affect the overall shot noise level, while higher densities would significantly increase the required integration times.

(Table 2). *Darwin* will thus provide statistically meaningful results on nearby planetary systems. K and M dwarfs are the easiest targets in terms of Earth-like planet detection for a given integration time, because on the one hand, the total thermal infrared luminosity of a planet in the Habitable Zone depends only on its size – while, on the other hand, the stellar luminosity is a strong function of its spectral type, so that the star/planet decreases for later spectral types. Compared to the case of the Sun and Earth, this contrast is two times higher for F stars, a factor of three lower for K stars, and more than an order of magnitude lower for M-dwarfs (Kaltenegger, 2008a). However, a special effort is placed on observing Solar-like G type stars (50% of the observing time) and a significant number of them can be screened and possible terrestrial planets studied.

**TABLE 1: EXPECTED PERFORMANCE OF THE X-ARRAY ARCHITECTURE IN TERMS OF NUMBER OF STARS SCREENED AND PLANETS CHARACTERISED FOR VARIOUS TELESCOPE DIAMETERS. ALL STARS ARE ASSUMED TO HOST AN EARTH-LIKE PLANET.**

|  | 1m diameter | 2m diameter | 3m diameter |
|---:|:---:|:---:|:---:|
| Screened | 76 | 218 | 405 |
| # F stars | 5 | 14 | 30 |
| # G stars | 15 | 53 | 100 |
| # K stars | 20 | 74 | 152 |
| # M stars | 36 | 77 | 123 |
| $CO_2$, $O_3$ spectroscopy | 17 | 49 | 87 |
| # F stars | 1 | 2 | 3 |
| # G stars | 4 | 8 | 15 |
| # K stars | 3 | 12 | 25 |
| # M stars | 9 | 27 | 44 |
| $H_2O$ spectroscopy | 14 | 24 | 43 |
| # F stars | 0 | 1 | 1 |
| # G stars | 2 | 4 | 7 |
| # K stars | 1 | 5 | 10 |
| # M stars | 11 | 14 | 25 |

If each nearby cool dwarf is surrounded by one rocky planet of one Earth radius within its Habitable Zone, only a fraction of the detected planets – about 25 of the most interesting planets - can be fully characterised (i.e., examined for the presence of $H_2O$, $CO_2$ and $O_3$) during the subsequent 3-year spectroscopic phase. This number would be doubled or quadrupled for planets with radii 1.5 and 2 times that of the Earth, respectively.

*Imaging for the General Astrophysics Program*

The 5σ, one hour, point source sensitivities for *Darwin* in 20% wide bands centred at 8, 10, 13 and 17 µm are approximately 0.1, 0.25, 0.5 and 0.8 µJy, respectively. These sensitivities are comparable to those of JWST. The maximum foreseen baselines are 500 metres, corresponding to a spatial resolution of 5 mas at 10 µm. Assuming a stability time scale of 10 seconds for the array, the sensitivity limit for self-fringe-tracking is about 10 mJy at 10 µm in a 0.5 arcsec aperture. This performance gives access to virtually all of the sources in the Spitzer SWIRE survey.

The importance of a bright source in the field to cophase the sub-pupils of the interferometer was mentioned earlier. The nature of the target and the science goal will determine the required instrumentation. We consider three different cases: few visibility measurements with a bright source in the Field of View, imaging with a bright source, and imaging without a bright source.

- With minimal impact on the nulling recombiner, *Darwin* can carry out visibility ($V^2$) science with JWST-like sensitivity, as long as there is a K ≤ 13 magnitude source in the field of view to stabilize the array. The modulus of the visibility provides simple size information about the target, for example, its radius assuming spherical morphology. The phase of the visibility gives shape information, such as deviations from spherical symmetry. If the target spectrum is smooth a few visibility measurements can be obtained rapidly, because *Darwin* can work simultaneously at several wavelengths. The baseline beam combiner could perform such measurements with very modest modification, and hence, with minimum impact on the cost of the mission. Therefore, a capability for basic visibility measurement should be implemented. Unfortunately, a limited number of $V^2$ observations provide useful information for only a limited number of targets.

- To obtain a fully reconstructed image, the (u,v) plane must be filled by moving the array. A significant gain in efficiency can be realized if the spectrum of the target is smooth over the operating band of the instrument. The shorter wavelengths sample higher spatial frequencies, and the longer wavelengths lower spatial frequencies, all at the same array spacing. A 100x100 image could be obtained with a few hundreds of positions, rather than the 10,000 positions required for a full spatial and spectral reconstruction.

- For targets with no bright source in the field, the preferred option for co-phasing is the use of a nearby off-axis bright reference star (K ≤ 13). One way of doing this is to feed the K-band light of this star along with the 6-20 µm light of the target to the Beam Combiner Satellite, a so-called dual field configuration. Another option is to make the interferometer optically rigid using Kilometric Optical Gyros. These devices can maintain the phasing of the array between pointing at a reference star and pointing at the target field. Clearly, this additional instrumentation may be much more demanding and expensive. A decision whether or not to add this capability to *Darwin* will depend on an analysis undertaken during the study phase and also available funding.

**TECHNOLOGY AND MISSION PLAN FOR *DARWIN***

*Essential Technology Developments for Darwin*

The pre-assessment study of *Darwin* by Alcatel in 2000, and the assessment study by TAS and Astrium in 2006 determined that there are no technology show stoppers for this ambitious mission. However, two key areas were identified that require focused attention and resources:

- *Formation Flying* of several spacecraft with relative position control of a few cm.
- The feasibility of *nulling interferometry* in the 6 - 20 µm range. Based on the expected star/planet contrast ($1.5 \times 10^{-7}$ at 10 µm and $10^{-6}$ at 18 µm for an Sun-Earth analogue) and on evaluations of instability noise, the common conclusion is that the null depth must be $10^{-5}$ on average, and that it must be sufficiently stable on the timescale of days so that the signal to noise ratio improves as the square root of time. This stability requirement translates into tight instrument control specifications, which can be relaxed by means of the two instability noise mitigation techniques (Lay, 2006; Lane *et al*., 2006). A thorough evaluation of these techniques and of the resulting instrumental stability requirements will be a key component of the technology development programme.

*Current Status of Technology Development*

Europe has devoted considerable resources, both intellectual and financial, to these technological issues since the initial Alcatel study (2000). ESA has invested approximately 20 M€ since 2000, with a significant increase in the last 2 years. Several tens of Technology Research Programs (TRPs) have been issued. NASA has run a parallel program in the US. Most of the key technologies have been addressed and significant progress achieved.

In the area of Formation Flying (FF), the TRPs "Interferometer Constellation Control" (ICC1 and ICC2) have developed nonlinear, high fidelity navigation simulators (Beugnon *et al*., 2004). Algorithms for Interferometer Constellation Deployment at L2 have also been demonstrated. In the USA, analogous simulations and a 2D robotic breadboard have shown the feasibility of formation flying (Scharf *et al*., 2007). Finally, with the PRISMA mission being prepared for launch, major hardware and software components of formation flying technologies will soon be demonstrated in space.

The investment in nulling interferometry research over the past 7 years has allowed the concept to be validated in European and American laboratories. The flight requirement is a null depth of $10^{-5}$ in the 6 – 20 µm domain. Monochromatic experiments using IR lasers at near-infrared and mid-infrared wavelengths have yielded nulls equal to or significantly better than $10^{-5}$ (Ergenzinger *et al*., 2004; Barillot *et al*., 2004; Buisset *et al*., 2006; Martin *et al*., 2007). Broadband experiments have achieved nulls of $1.2 \times 10^{-5}$ for 32% bandwidth at 10 µm, closely approaching the flight requirement (Peters *et al*., 2006). At the time of writing the technology of nulling interferometry is nearing maturity, although it has not yet been demonstrated over the full *Darwin* bandwidth with the required depth and stability.

Additional key technological developments in recent years include:

- Selection of the baseline *interferometer configuration*. Significant effort in this area since 2000, backed by independent studies in the US and Europe, has identified the non-planar Emma X-Array as the optimal choice (Lay, 2005; Karlsson *et al*., 2006);

- *Achromatic Phase Shifters (APS),* which allow broadband destructive interference between beams, have been demonstrated in the laboratory. A comparative study currently running in Europe should identify the preferred approach (Labèque *et al*., 2004);

- *Space-qualified Delay Lines* to balance the different optical paths to nanometre accuracy have been demonstrated (Van den Dool *et al*., 2006). A breadboard at TNO-TPD has achieved this performance at 40 K and may be included as a test payload in the PROBA 3 space mission;

- *Single Mode Fibres*, or *Integrated Optics Modal Filters* that enable broadband nulling have recently been produced and tested (Labadie *et al*., 2006; Huizot *et al*., 2007). Chalcogenide fibres have demonstrated the required performance of 40% throughput and 30 dB rejection of higher order spatial modes. Ongoing work is emphasizing silver halide single-mode filters, which will operate in the 12-20 µm band (Lewi *et al*., 2007). Photonic Crystal fibres that can cover the whole spectral domain in a single optical channel are considered (Flanagan *et al*., 2006);

- *Detector Arrays* with appropriate read noise and dark current have been qualified for space-based operation (Love *et al*., 2004). The Si:As Impurity Band Conductor (IBC) arrays developed for JWST appear to be fully compliant with *Darwin* requirements. A reduced-size version of the JWST 1024 x 1024 detector, e.g. 512 x 8 (300 x 300 for the general astrophysics program), could be read out at the required rate with a dissipation of a few tens to hundreds of µW. These devices exhibit high quantum efficiency (80%), low read noise (19 $e^-$), and minimal dark current (0.03 $e^-$/s at 6.7 K). Such performance permits sensitive observations, even at moderately high spectral resolution ($R = 300$);

- Low vibration *Cryo-coolers* for the detector system have been demonstrated in the laboratory. A European TRP has led to a prototype absorption cooler providing 5 mW of cooling power at 4.5 K (Burger *et al*., 2003). JPL scientists have demonstrated a system with 30 mW of cooling at 6 K (Ross and Johnson, 2006).

**Conclusions:**

We described the scientific and technological implementation of the *Darwin* space mission, whose primary goal is the search and characterization of terrestrial extrasolar planets as well as the search for life. *Darwin* is designed to detect and characterize rocky planets similar to the Earth and perform spectroscopic analysis of their atmosphere at mid-infrared wavelengths (6 to 20 µm). The baseline mission lasts 5 years and consists of approximately 200 individual target stars. Among these, 25 to 50 planetary systems can be studied spectroscopically, searching for gases such as $CO_2$, $H_2O$, $CH_4$ and $O_3$. Key technologies required for the construction of *Darwin* have already been demonstrated. This paper has described the science programme and some of the technological requirements for an ambitious space mission to discover and characterize Earth-like planets and to search for evidence of life on them. The *Darwin* mission will address one of the most fundamental questions: humankind's origin, and its place in the Universe.

**FIGURES**

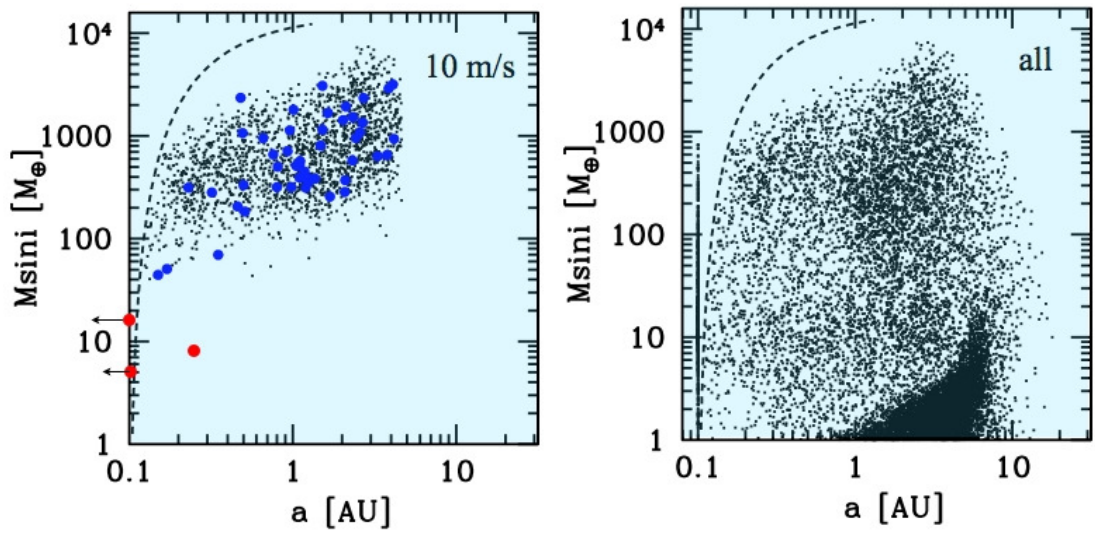

Figure 1. An example of synthetic planet populations, generated by computation we made, which allow a statistical comparison with observations. Left : black points are the giant planets predicted by the model, the circles are the giant planets actually detected, in reasonable agreement with predictions. Right: prediction of the same model for small planets (e.g. terrestrial) which are not yet detectable. It appears that many Earth-mass or somewhat bigger planets are expected which would be interesting targets for *Darwin*.

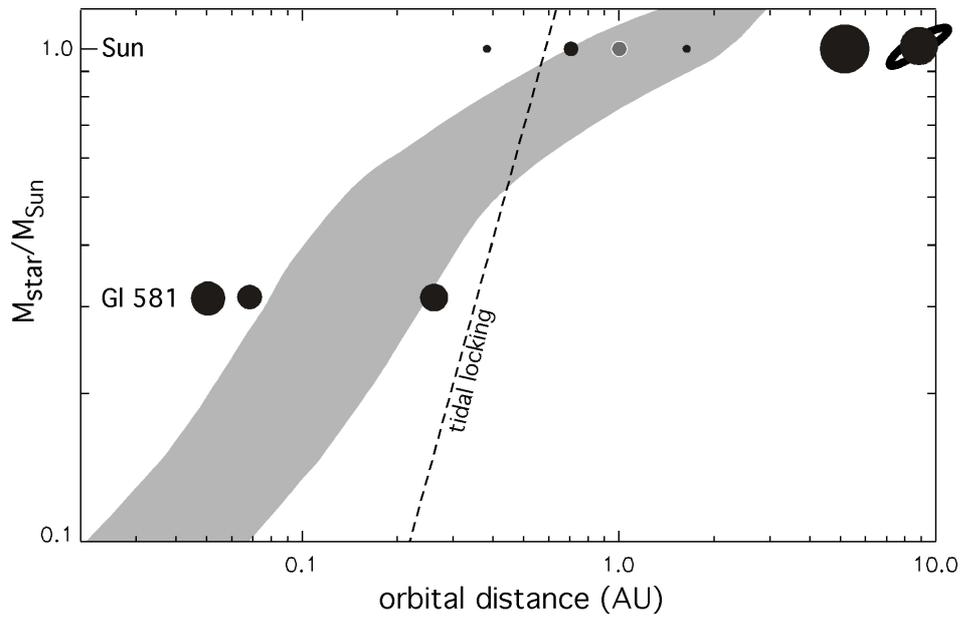

Figure 2. The limits of the Habitable Zone (grey box) as a function of the stellar mass.

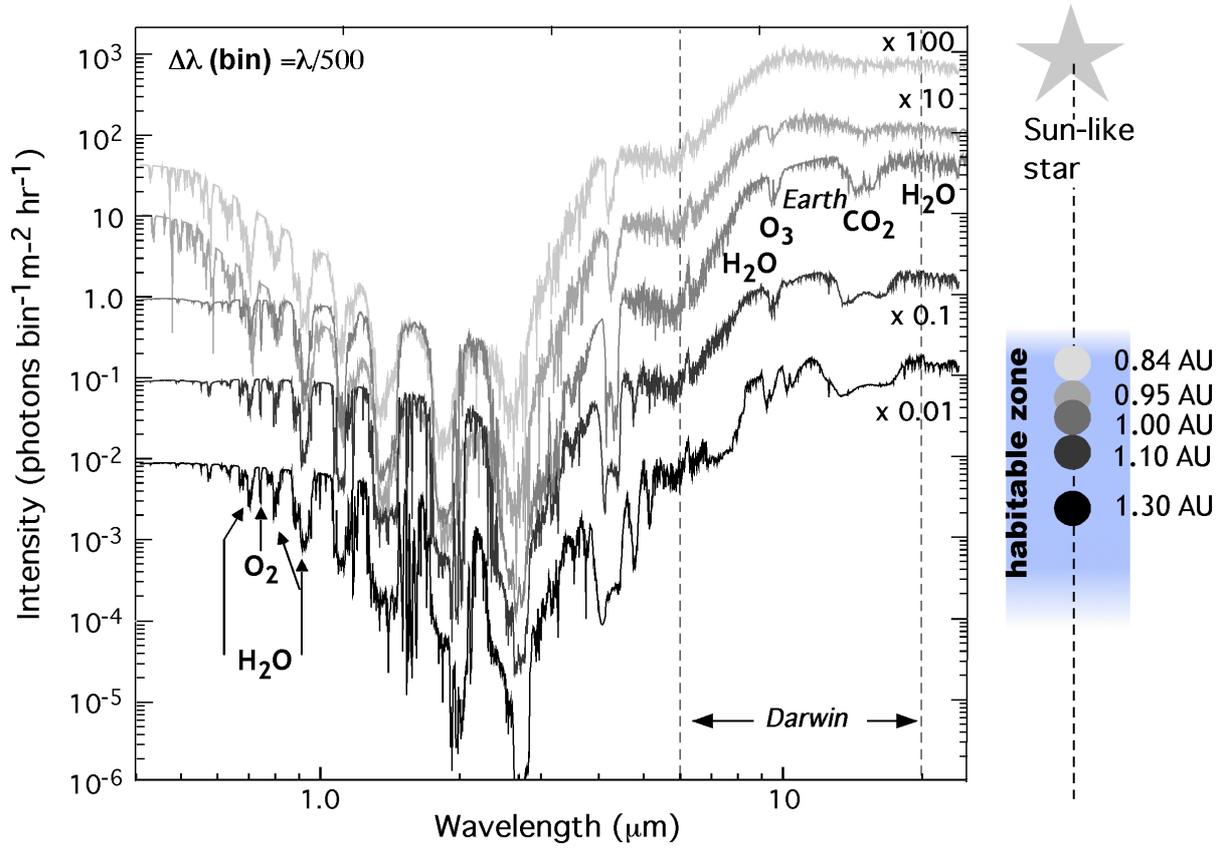

Figure 3. The estimated evolution of the $H_2O$, $O_3$ and $CO_2$ features in the spectra of an Earth-like planet as a function of its location in the HZ. Sun-like star refers to a 1 solar mass G star.

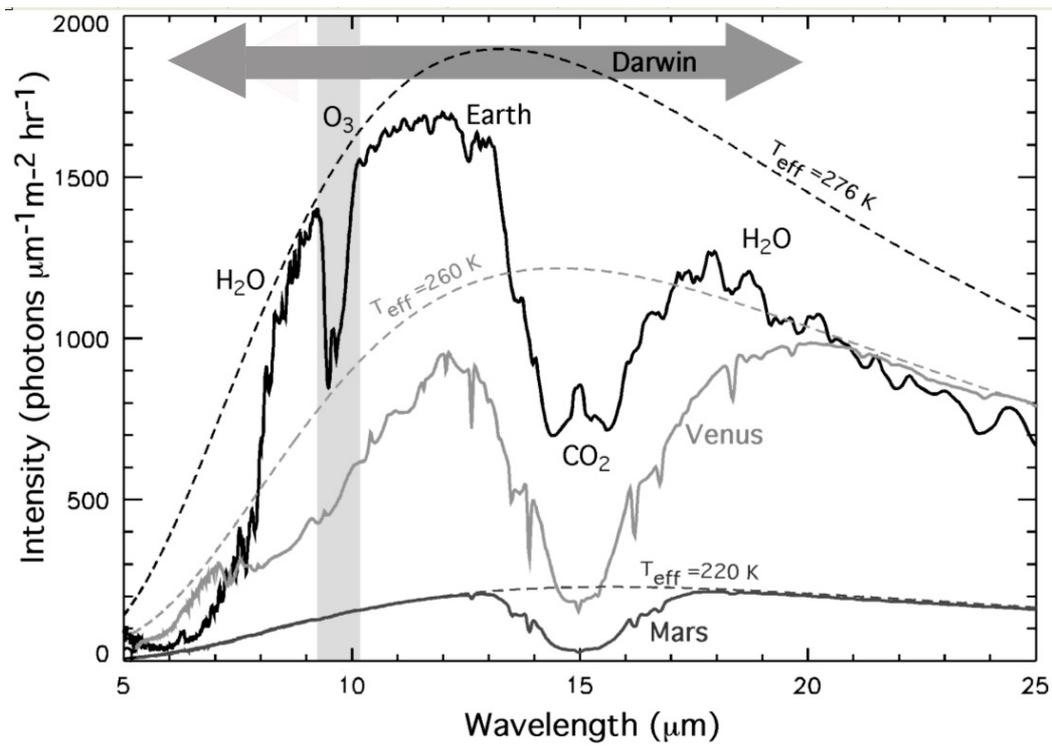

Figure 4. The mid-IR spectrum of the Earth, Venus and Mars at a low resolution (spectra are derived from a variety of published models including Meadows and Crisp, 1996; Tinetti *et al.*, 2005; Tinetti *et al.*, 2006; Kaltenegger *et al.*, 2007; Selsis *et al.*, 2007b).

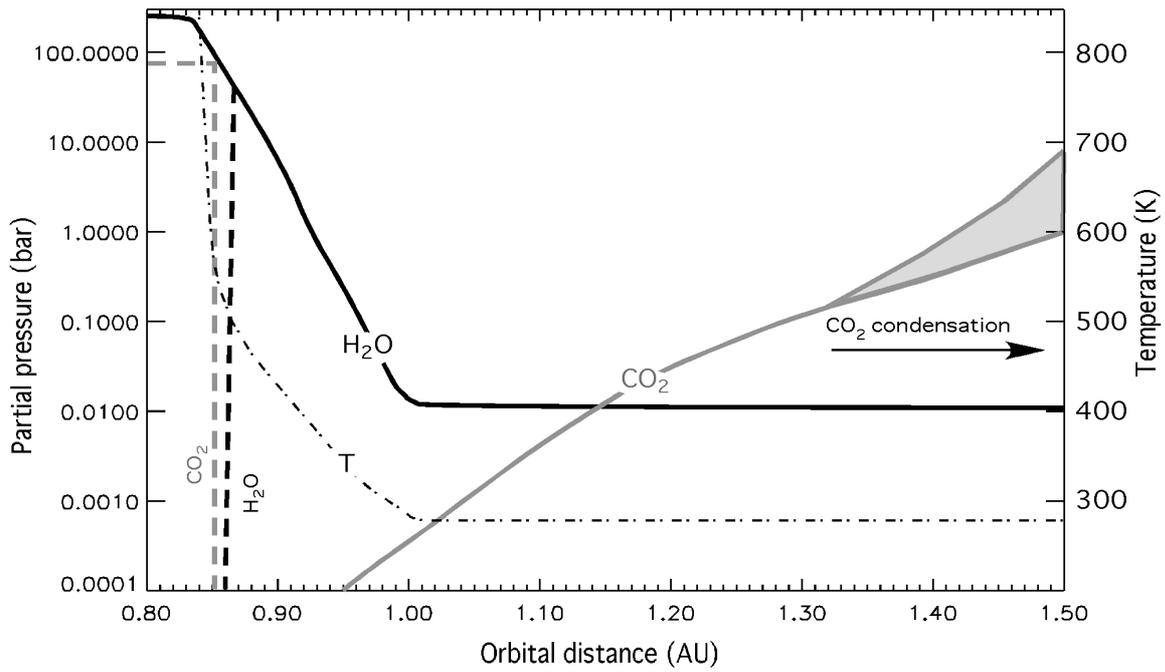

Fig. 5. Diagram illustrating the reason for the spectra shown in Figure 4. The mean surface temperature (T) and partial pressure of $CO_2$ and $H_2O$ as a function of the orbital distance on a habitable planet within the habitable zone (Kaltenegger and Selsis, 2007). Data adapted from Kasting *et al*. (1993) and Forget and Pierehumbert (1997). (Partial pressure, left y-axis and Surface Temperature, right y-axis).

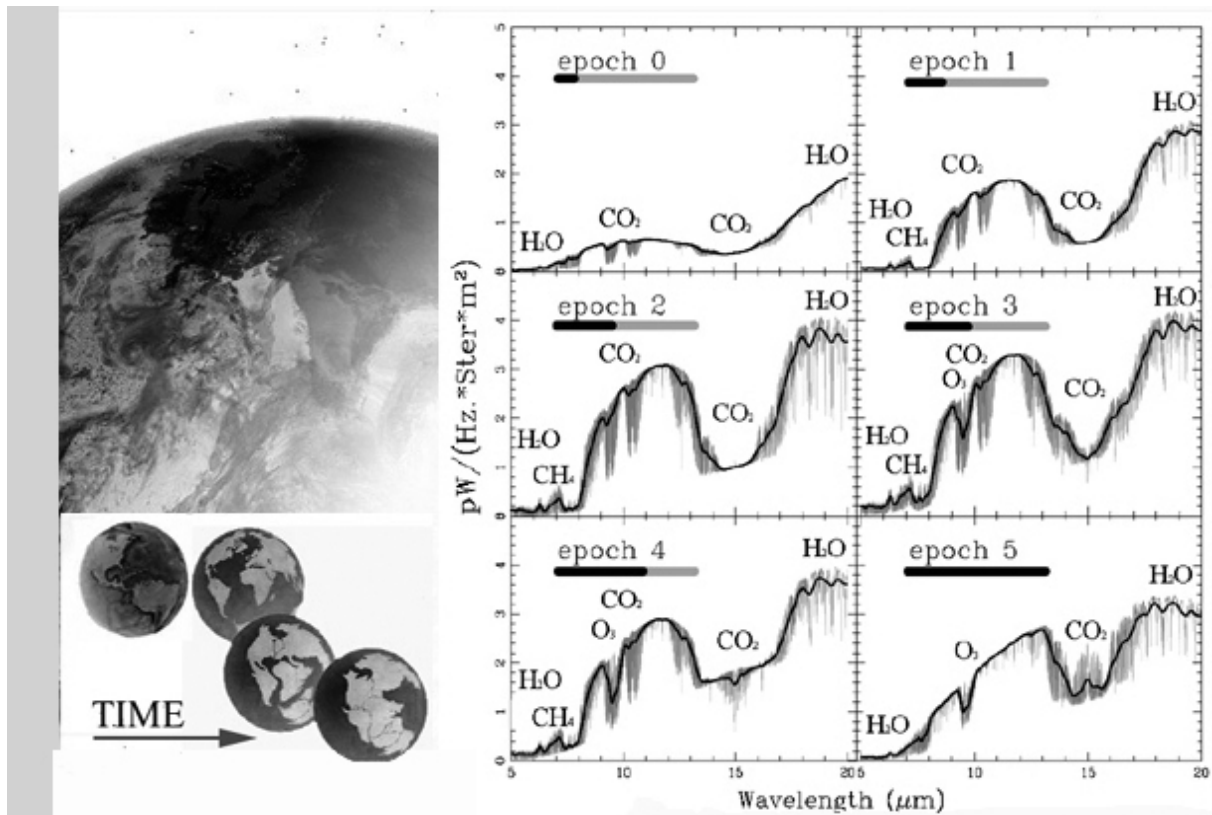

Figure 6. Mid-IR synthetic spectra of the Earth at six different stages of its evolution: 3.9 (Epoch 0), 3.0, 2.6, 2.0, 0.8 (Epoch 4) Gyrs ago and the present-day Earth (Epoch 5) (figure from Kaltenegger *et al*., 2007)

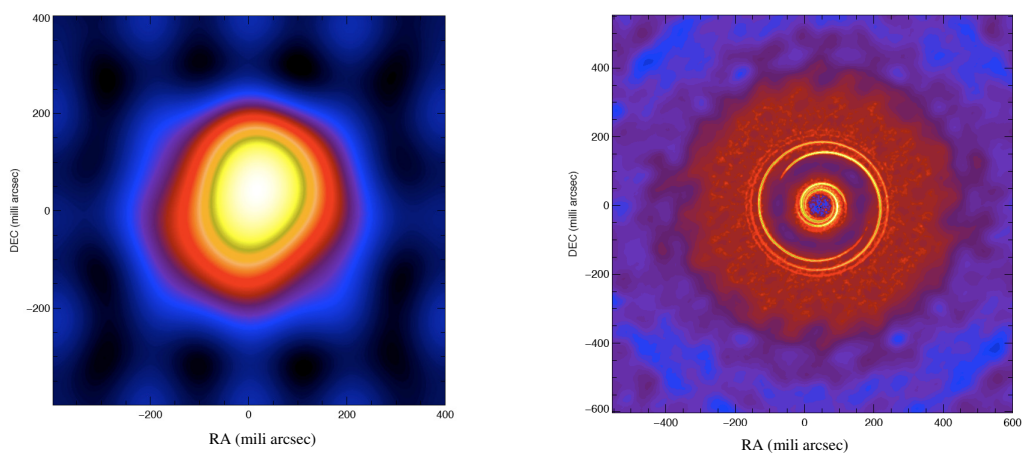

Figure 7. An illustration of the mapping power of the mission. Simulation of a hot accretion disk in the Taurus cloud (140 pc) as seen by JWST (Left) and *Darwin* in its imaging mode (Right). Simulated JWST and *Darwin* images are based on scaled models by D'Angelo et al. (2006) for the formation of a planet of one Jupiter mass at 5.2 AU, orbiting a solar type star. The most prominent feature in the model is a low-density annular region along the planet's path (known as gap) and spiral wave patterns. Total observing time is 10 h. (courtesy Cor de Vries)

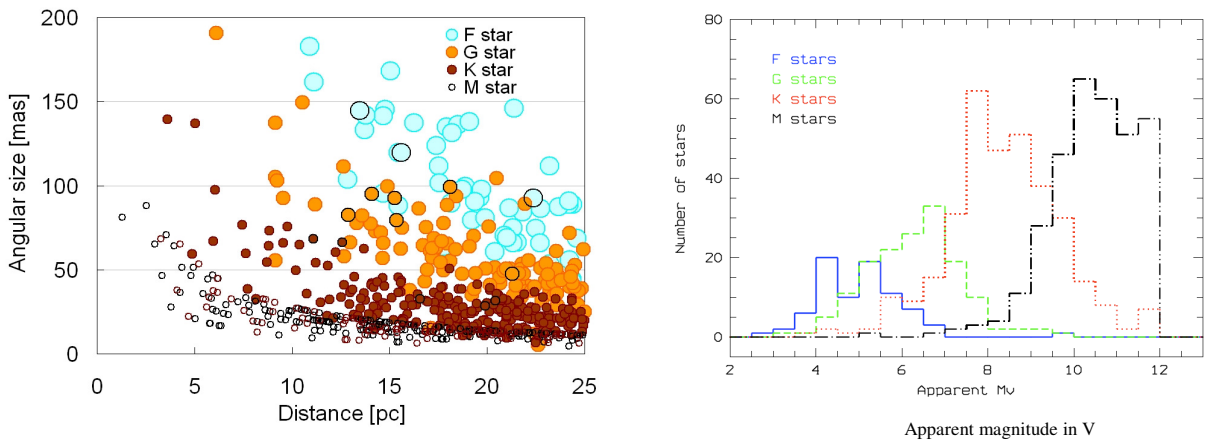

Figure 8. Some features of stars in the *Darwin* star catalog. Left, size of the Habitable Zone for the different spectral types of *Darwin* targets (Kaltenegger et al, 2008a). Right, histogram of their apparent visible magnitudes.

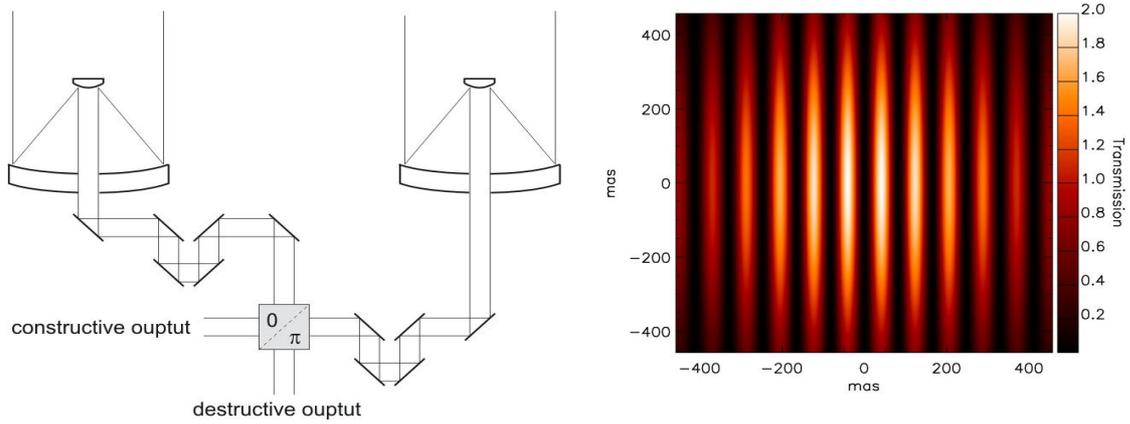

Figure 9. The concept of nulling interferometry. Left, principle of a two-telescope Bracewell nulling interferometer with its destructive (null) and constructive outputs. Right, associated transmission map, displayed for $\lambda = 10$ µm and a 25-m baseline array. This fringe pattern is effectively projected on the sky, blocking some regions, e.g. the on-axis star, while transmitting others, e.g. the off-axis planet.

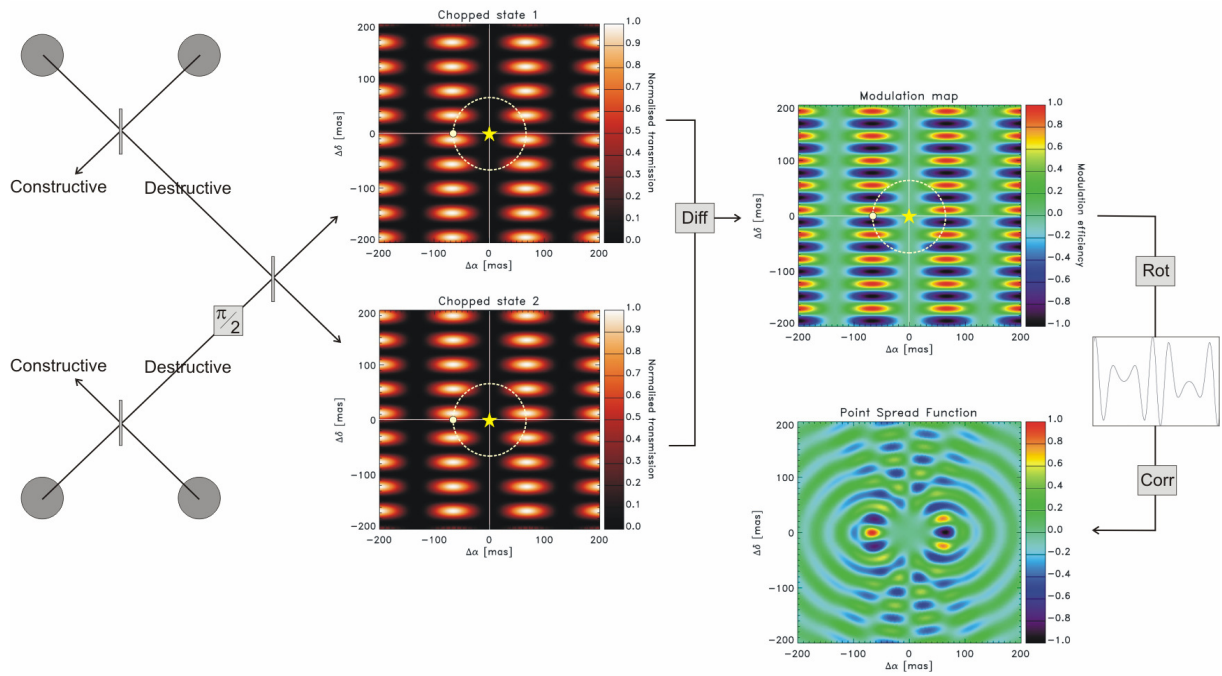

Figure 10. Phase chopping for the X-array, a four-element rectangular configuration of telescopes. Combining the beams with different phases produces two conjugated chopped states, which are used to extract the planetary signal from the background. Array rotation then locates the planet by cross-correlation of the modulated chopped signal with a template.

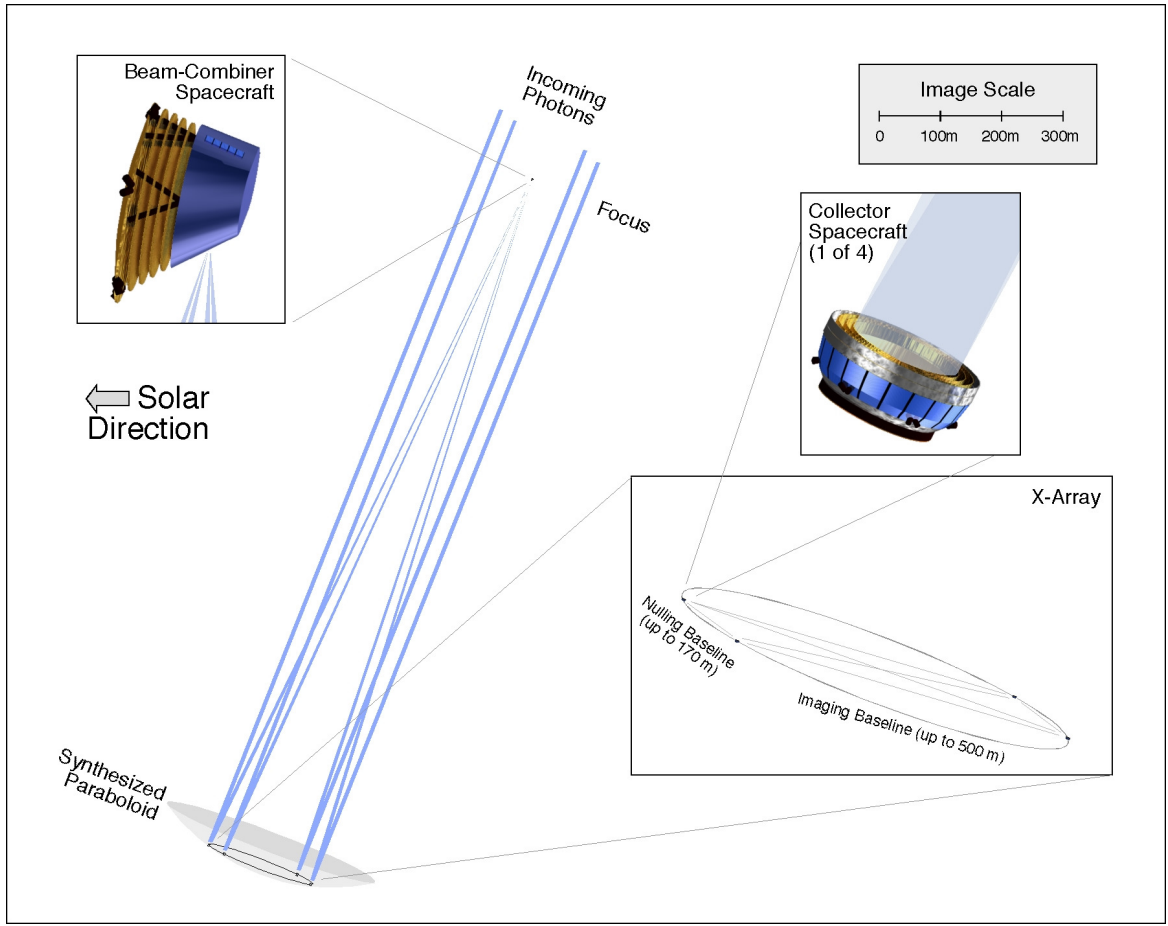

Figure 11. The non-coplanar Emma X-array configuration. It consists of 4 collector spacecrafts and a beam combiner spacecraft. Spherical mirrors in the collectors form part of a large, synthetic paraboloid, feeding light to the beam combiner at its focus.

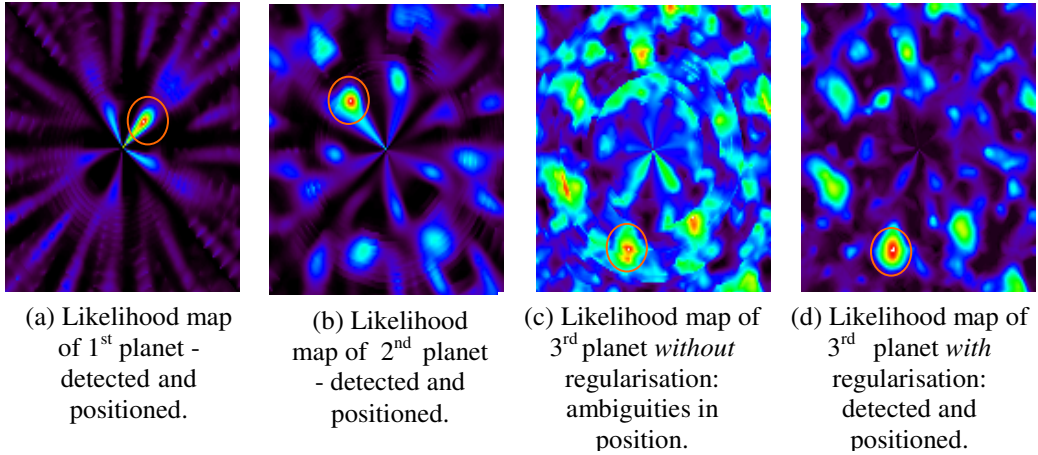

(a) Likelihood map of 1st planet - detected and positioned.

(b) Likelihood map of 2nd planet - detected and positioned.

(c) Likelihood map of 3rd planet *without* regularisation: ambiguities in position.

(d) Likelihood map of 3rd planet *with* regularisation: detected and positioned.

Figure 12. Example of the Bayesian approach to image reconstruction. Likelihood maps of the successive detection of 3 planets located at 0.64, 1.1 and 1.8 AU of a star. Red indicates a higher probability, black a lower one, white is the highest. The spectral resolution is 15 and S/N is 0.33 per spectral element. The 3rd and faintest planet is correctly detected when the so-called regulation process is used [compare (c) and (d)]